\documentclass[longauth]{aa} 
\usepackage{lineno}
\usepackage{graphicx}
\usepackage{txfonts}
\usepackage{color}
\usepackage{natbib}
\usepackage{tabularx}
\usepackage{multirow}
\usepackage{url}

\def\degr{\hbox{$^\circ$}}
\def\arcmin{\hbox{$^\prime$}}
\def\arcsec{\hbox{$^{\prime\prime}$}}

\begin{document}
%


\title{Constraints on the gamma-ray emission from the cluster-scale AGN outburst in the Hydra~A galaxy cluster}

\author{HESS Collaboration
\and A.~Abramowski \inst{1}
\and F.~Acero \inst{2}
\and F.~Aharonian \inst{3,4,5}
\and A.G.~Akhperjanian \inst{6,5}
\and G.~Anton \inst{7}
\and S.~Balenderan \inst{8}
\and A.~Balzer \inst{7}
\and A.~Barnacka \inst{9,10}
\and Y.~Becherini \inst{11,12}
\and J.~Becker \inst{13}
% ...
}

\institute{
Universit\"at Hamburg, Institut f\"ur Experimentalphysik, Luruper Chaussee 149, D 22761 Hamburg, Germany \and
Laboratoire Univers et Particules de Montpellier, Universit\'e Montpellier 2, CNRS/IN2P3,  CC 72, Place Eug\`ene Bataillon, F-34095 Montpellier Cedex 5, France \and
Max-Planck-Institut f\"ur Kernphysik, P.O. Box 103980, D 69029 Heidelberg, Germany \and
Dublin Institute for Advanced Studies, 31 Fitzwilliam Place, Dublin 2, Ireland \and
National Academy of Sciences of the Republic of Armenia, Yerevan  \and
Yerevan Physics Institute, 2 Alikhanian Brothers St., 375036 Yerevan, Armenia \and
Universit\"at Erlangen-N\"urnberg, Physikalisches Institut, Erwin-Rommel-Str. 1, D 91058 Erlangen, Germany \and
University of Durham, Department of Physics, South Road, Durham DH1 3LE, U.K. \and
Nicolaus Copernicus Astronomical Center, ul. Bartycka 18, 00-716 Warsaw, Poland \and
CEA Saclay, DSM/IRFU, F-91191 Gif-Sur-Yvette Cedex, France \and
APC, AstroParticule et Cosmologie, Universit\'{e} Paris Diderot, CNRS/ IN2P3,CEA/ lrfu, Observatoire de Paris, Sorbonne Paris Cit\'{e}, 10, rue Alice Domon et L\'{e}onie Duquet, 75205 Paris Cedex 13, France,  \and
Laboratoire Leprince-Ringuet, Ecole Polytechnique, CNRS/IN2P3, F-91128 Palaiseau, France \and
Institut f\"ur Theoretische Physik, Lehrstuhl IV: Weltraum und Astrophysik, Ruhr-Universit\"at Bochum, D 44780 Bochum, Germany
% ...
}

\offprints{\email{wilfried.domainko@mpi-hd.mpg.de}} 

\date{}
 
\abstract
{In some galaxy clusters powerful AGN have blown bubbles with cluster scale extent into the ambient medium. The main pressure support of these bubbles is not known to date, but cosmic rays are a viable possibility. For such a scenario copious gamma-ray emission is expected as a tracer of cosmic rays from these systems.}
{Hydra~A, the closest galaxy cluster hosting a cluster scale AGN outburst, located at a redshift of 0.0538, is investigated for being a gamma-ray emitter with the High Energy Stereoscopic System (H.E.S.S.) array and the \emph{Fermi} Large Area Telescope (\emph{Fermi}-LAT).}
{Data obtained in 20.2 hours of dedicated H.E.S.S. observations and 38 months of \emph{Fermi}-LAT data, gathered by its usual all-sky scanning mode, have been analyzed to search for a gamma-ray signal.}
{No signal has been found in either data set. Upper limits on the gamma-ray flux are derived and are compared to models. These are the first limits on gamma-ray emission ever presented for galaxy clusters hosting cluster scale AGN outbursts.}
{The non-detection of Hydra~A in gamma-rays has important implications on the particle populations and physical conditions inside the bubbles in this system. For the case of bubbles mainly supported by hadronic cosmic rays, the most favorable scenario, that involves full mixing between cosmic rays and embedding medium, can be excluded. However, hadronic cosmic rays still remain a viable pressure support agent to sustain the bubbles against the thermal pressure of the ambient medium. The largest population of highly-energetic electrons which are relevant for inverse-Compton gamma-ray production is found in the youngest inner lobes of Hydra~A. The limit on the inverse-Compton gamma-ray flux excludes a magnetic field below half of the equipartition value of 16~$\mu$G in the inner lobes.}

\titlerunning{Gamma-ray observations of Hydra~A}
\authorrunning{HESS collaboration}

\maketitle

\section{Introduction}

At the center of some galaxy clusters powerful Active Galactic Nuclei (AGN) reside and the feedback of outbursts generated by these AGN on the embedding Intra-Cluster Medium (ICM) can be seen in several systems \cite[for a review, see][]{mcnamara2007}. Typical signatures for an AGN -- ICM interaction are surface brightness depressions in the diffuse thermal X-ray emission of the cluster which are caused by cavities in the ICM. These cavities appear to be filled with non-thermal electrons which radiate in the radio band due to synchrotron emission \citep[e.g.][]{birzan2004,dunn2006}. AGN-blown bubbles surrounded by thermal plasma offer the exciting possibility to constrain the energetics of these outbursts. This can be done by estimating the work that is necessary to expand the bubbles against the thermal pressure of the embedding ICM (pV work in the following). The energetics involved in this AGN activity can be enormous, in some cases even exceeding 10$^{61}$~erg \citep{mcnamara2007}. The most powerful AGN outbursts known to date are found in MS~0735+7421 \citep{mcnamara2005}, Hercules~A \citep{nulsen2005a} and Hydra~A \citep{nulsen2005b}. The AGN created bubbles in these systems have ages of about 10$^8$ years and exhibit sizes on the scale of the galaxy cluster itself. 

The nature of the main pressure support agent which fills the bubbles in the ICM is not known to date. Viable possibilities for the pressure support in such systems would be relativistic particles such as hadronic cosmic rays
or electrons \citep[e.g.][]{dunn2004,ostrowski2001,hinton2007}, magnetic fields \citep[e.g.][]{dunn2004} or hot plasma \citep[e.g.][]{gitti2007}. The energy required to expand bubbles with volume $V$ into a surrounding ICM with pressure $p$ ranges from $2pV$ for magnetic fields to $4pV$ for relativistic fluids such as cosmic rays \citep[e.g.][]{wise2007}.

One inevitable consequence of bubbles filled with non-thermal particles would be the production of gamma-ray emission. For the case of hadronic cosmic rays, gamma rays are produced by inelastic collisions between the high energy particles and the thermal surrounding medium \citep[e.g.][]{hinton2007}. In case of electrons, gamma-rays are produced by up-scattering of cosmic microwave background (CMB) and infrared extragalactic background light (EBL) photons by these electrons \citep[e.g.][]{abdo2010a}. Large-scale leptonic gamma-ray emission connected to AGN lobes has indeed been discovered with the \emph{Fermi} satellite from the radio galaxy Centaurus~A \citep{abdo2010a} and potentially from NGC~6251 \citep{takeuchi2012}. Both galaxies are not hosted by a cluster. To date, no galaxy cluster has been firmly detected in gamma rays \citep{perkins2006,aharonian2009a,aharonian2009b,aleksic2010,ackermann2010}. 
However, the detection of an extended  gamma-ray signal resulting from annihilation emission from supersymmetric dark matter has been claimed for the Virgo, Fornax and Coma cluster \citep{han2012}. Recently, NGC~1275 the central radio galaxy of the Perseus cluster has been detected in VHE gamma rays \citep{aleksic2012a}, and with this deep exposure stringent upper limits on the emission of the Perseus cluster itself have been obtained \citep{aleksic2012b}.
Galaxy clusters hosting cluster-scale AGN outbursts appear to be promising targets for gamma-ray observations according to the extraordinary energetics inferred from the AGN -- ICM interaction seen in these systems. To date, no gamma-ray observations on galaxy clusters that host cluster-scale AGN outbursts have been presented.

The Hydra~A system (Abell~0780) at RA(J2000)~=~9$^\mathrm{h}$18$^\mathrm{m}$05.7$^\mathrm{s}$ and 
Dec(J2000)~=~-12\degr05\arcmin44\arcsec at a redshift of 0.0538 is the closest known galaxy cluster which hosts a cluster-scale AGN outburst \citep{nulsen2005b}. It features several cavities with a total expansion work $pV$ of $4 \times 10^{60}$~erg done on the ICM. Thus the total energy required, depending on the equation of state of the main pressure agent, is $(0.8 - 1.6) \times 10^{61}$ erg which were deposited in the last few 10$^8$ years in the surroundings \citep{wise2007}. Hydra~A also features low-frequency radio lobes extending to almost 4\arcmin\ from the cluster center \citep{lane2004}. Shocks in the ICM surround these radio lobes \citep{nulsen2005b} with energetics of $9 \times 10^{60}$ erg, comparable to the expansion work done in the cavities against the thermal plasma. The central AGN outburst has also driven substantial gas dredge-up in the Hydra~A system \citep{gitti2012}.
 
\emph{Chandra} has furthermore revealed an extensive cavity system consisting of three generations of cavities with decreasing ages, which points towards a complex activity history of the system \citep{wise2007}. Most relevant for gamma-ray production are the giant outer lobes that dominate the energetics in the Hydra~A system and the inner lobes that are expected to contain the youngest population of particles. Both possibilities will be further discussed in Sec.~\ref{sec:discussion}.

Due to its proximity and energetics, Hydra~A is expected to feature the highest gamma-ray flux of all galaxy clusters harboring cluster-scale AGN outbursts.
For the case of hadron-dominated bubbles it was inferred that the flux might be close to the detection limit of the current generation of gamma-ray instruments \citep{hinton2007}.

In this paper, upper limits on the gamma-ray emission from the Hydra~A system are reported. Limits obtained by the High Energy Stereoscopic System (H.E.S.S.) and the \emph{Fermi} Large Area Telescope (\emph{Fermi}-LAT) are presented in Sec.~\ref{sec:hess} and Sec.~\ref{sec:fermi}, respectively.
These limits are used to obtain constraints on the energy of hadronic cosmic rays (Sec.~\ref{sec:hadronic}) and electrons (Sec.~\ref{sec:leptonic} and \ref{sec:sync-ic}) which may populate the AGN outburst region in this galaxy cluster.

Throughout this paper a $\Lambda$CDM cosmology with $H_0$~=~70~km~s$^{-1}$~Mpc$^{-1}$, $\Omega_{\Lambda}$ = 0.7 and  $\Omega_{M}$ = 0.3 is assumed, corresponding to a luminosity distance of $d_\mathrm{L}$ = 240 Mpc, an angular diameter distance of 216 Mpc and a linear scale of 1.05 kpc per arcsecond \citep{wise2007}.


\section{H.E.S.S. data analysis}
\label{sec:hess}

Hydra~A was observed in the VHE gamma-ray range (E~$>$~100~GeV) with H.E.S.S. \citep{hinton2004}, which is an array of four Imaging Atmospheric Cherenkov Telescopes located at the Khomas Highland in Namibia (23$^\circ$16'18'' S 16$^\circ$30'00'' E, altitude 1800 m). Data were taken in March and April 2007 and from January to March 2010. In total 20.2 hours of good quality data \citep[excluding data taken during bad weather and data affected by hardware irregularities, see][]{aharonian2006} were collected. The data were obtained with a mean zenith angle of 15$^\circ$ which resulted in an energy threshold of 240~GeV.

\begin{figure}[ht]
\centering
\includegraphics[width=8cm]{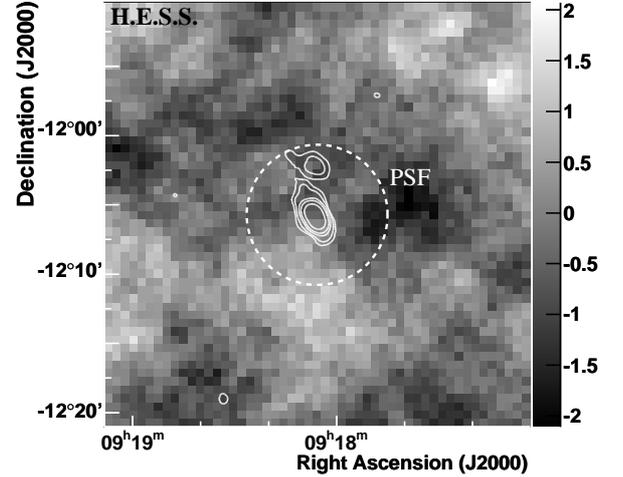}

\caption{Significance map from the H.E.S.S. data. Overlaid are contours of the NVSS 1.4 GHz radio survey \citep{condon1998} which show the extension of the lobes in Hydra~A. Additionally the 68\% containment radius of the H.E.S.S. PSF is indicated by the dashed circle.} 

\label{figure:excess_hess}
\end{figure}

The data were analyzed with a boosted decision tree method \citep{ohm2009}. 
For H.E.S.S., Hydra~A was treated as a point-like source. This is a reasonable assumption since
the lobes of Hydra~A extend over 4\arcmin\, in comparison to the 68\% containment radius of the H.E.S.S. point spread function (PSF) of 6\arcmin\ \citep{aharonian2006}. Using $\zeta$ {\tt std} cuts \citep[$\zeta$ denotes the boosted decision tree classifier and for the definition of {\tt std cuts} see][]{ohm2009}\footnote{software version hap-11-02-pl07} and {\tt reflected} background model \citep{berge2007}, a total number of counts $N_\mathrm{ON}$ of 456 on the target and a number of background counts $N_\mathrm{OFF}$ of 7265 (with source to background normalization $\alpha = 0.0614$) were measured. This results in an excess of 9.7 events corresponding to a significance of 0.4$\sigma$ \citep{li1983} and hence no significant signal has been found (see Fig.~\ref{figure:excess_hess}). This result was confirmed with an independent calibration and analysis chain \citep{denaurois2009}. Since no signal was detected, upper limits were derived using the method of \citet{rolke2005}. At a confidence level of 95\% an upper limit of $F_\gamma(>240\, \mathrm{GeV}) < 7.9 \times 10^{-13}$~cm$^{-2}$s$^{-1}$ for a power-law of the form $dN/dE \propto E^{-\Gamma}$ with an assumed photon index $\Gamma = 2.5$ is found. The gamma-ray index is chosen to approximately match the shape of the predicted spectrum \citep[see][]{hinton2007}.
Upper limits were computed also for $\Gamma = 2.0$ and only weakly depend on the spectral index with a difference of less than 10\%. In Fig.~\ref{figure:sed} upper limits for $\Gamma = 2.5$ that fits closest to the model predictions, are plotted and compared to model predictions.

\section{\emph{Fermi} data analysis}
\label{sec:fermi}

Hydra~A has been observed by the  Large Area Telescope (LAT), which is the primary instrument on the \emph{Fermi} Gamma-ray Space Telescope (\emph{Fermi}).
It is a pair conversion telescope for high-energy gamma-rays with a wide field of view. It covers the energy regime from 20 MeV to 300 GeV with an angular resolution of approximately 3.5$^\circ$ at 100 MeV and narrowing to 0.14$^\circ$ at 10 GeV \citep[see][]{atwood2009}. In survey mode, the observatory is rocked north and south on alternate orbits so that every part of the sky is observed for $\sim$30 minutes every 3 hours. 

\emph{Fermi}-LAT observations of Hydra A from MJD 54682.9 to 55816.7, corresponding to a period of $\sim$38 months from August 2008 to September 2011, are used in this paper. The data were retrieved from the public data archive and analyzed using the \emph{Fermi} Science Tools v9r23 package. The standard event filtering, reconstruction and classification were applied to the data \citep{abdo2009}. The instrument response function, {\tt P7SOURCE\_V6}, is applied throughout the data analysis.

Events with energies between 200 MeV and 200 GeV and within a circular region of 15$^\circ$ radius have been considered in the analysis. A binned maximum likelihood analysis was performed on the data using the {\tt gtlike} tool. All sources in the \emph{Fermi} second year catalog \citep{abdo2011} within 15$^\circ$ are included in the source model as well as the Galactic diffuse emission model, {\tt gal\_2yearp7v6\_v0.fits}, and the corresponding extragalactic isotropic diffuse emission model. The isotropic model used is {\tt iso\_p7v6source.txt}, which is valid for {\tt P7SOURCE\_V6} instrument response functions. All point sources were modeled with parameters fixed to those reported in the \emph{Fermi} second year catalog unless they were within 10$^\circ$ of Hydra~A. Since Hydra~A does not appear in the \emph{Fermi} second year catalog, an additional point source was inserted at its position, assuming a power-law spectrum. Details on the likelihood analysis techniques and the models used can be found on the \emph{Fermi} Science Support Center website\footnote{ \url{http://fermi.gsfc.nasa.gov/ssc/}} \citep[see also][]{abdo2009}. 

The TS value, which square root is a measure of the significance of a source, at the position of Hydra~A is about 1. This implies that no significant signal has been found in the data.
Therefore, 95\% flux upper limits are produced for a point-like source with an assumed spectral index of $\Gamma$ (1.5, 2.0 and 2.5, that are given in Table~\ref{table:upper_limits_fermi}). Models are compared to upper limits obtained for the spectral index that is most compatible with the prediction. Flux upper limits are shown in Fig.~\ref{figure:sed} for $\Gamma = 1.5$ (blue) and $\Gamma = 2.0$ (cyan) whereas the limit obtained for $\Gamma = 2.5$ is shown in Fig~\ref{figure:cena} and \ref{F-SED}. 

\begin{table}
\begin{center}
\begin{tabular}{l|c}
\hline
\hline
$\Gamma$ & $F_\mathrm{ul}$($>$200\,MeV) [erg cm$^{-2}$ s$^{-1}$]\\ 
\hline
1.5 & $1.4 \times 10^{-12}$ \\
2.0 & $2.1 \times 10^{-13}$ \\
2.5 & $3.2 \times 10^{-13}$ \\
\hline
\end{tabular}
\caption{Fermi-LAT upper limits obtained for the AGN outburst in Hydra~A.} \label{table:upper_limits_fermi}
\end{center}

\end{table}

\section{Discussion}
\label{sec:discussion}

Several scenarios of gamma-ray production in the Hydra~A galaxy cluster are possible. The whole system consists of three generations of AGN outbursts that created bubbles of different energetics and consecutive ages \citep[see][]{wise2007}. The giant outer lobes inflated by the oldest cycle of AGN activity contain most of the energy. For these outer lobes a hadronic scenario is investigated in Sec.~\ref{sec:hadronic} and a potential leptonic scenario is discussed in Sec.~\ref{sec:leptonic}. In a hadronic scenario it is expected that gamma-ray emission is dominated by the giant outer lobes, since cooling of hadronic cosmic rays on timescales relevant for the evolution of the outer lobes of $\sim10^{8}$ years \citep[see][]{hinton2007} is unimportant in the Hydra~A system. The situation is different for electrons since they are effected by cooling on this rather long time scale. It is expected that the largest population of highly-energetic electrons which are relevant for IC gamma-ray production is found in the youngest inner lobes. Therefore, a leptonic scenario for the inner lobes is examined in Sec.~\ref{sec:sync-ic}.

\subsection{Constraints on the particle population in the giant outer lobes}

\subsubsection{Hadronic population}
\label{sec:hadronic}

The gamma-ray luminosity in a hadronic scenario is given by the total energy in cosmic rays and the mean density of target material. In galaxy clusters which harbor cluster-scale AGN outbursts, the total energy in cosmic rays can in principle be estimated from the energetics of the AGN outbursts if this is assumed to be the energetically dominant feedback agent on the ICM. Viable proxies for the energy in cosmic rays could be the energetics of the shock wave or the energy needed to sustain the X-ray cavities. In the general picture for AGN outbursts in galaxy clusters, the radio bubbles are dominated by cosmic rays, whereas the thermal ICM is distributed around these bubbles. The radio bubbles appear as surface brightness depressions in X-rays. This indicates a depletion of the hot ICM inside them. The three-dimensional structure of the bubbles is not known and, consequently, the actual density of X-ray emitting plasma inside the cavities can only loosely be constrained. Limits on the density of thermal plasma inside the lobes can also be obtained with the depolarization effect of the radio emission of the lobes \citep{garrington1991}. The actual gamma-ray luminosity of the system will depend on the level of mixing between hadronic cosmic rays and the thermal ICM. Processes which can lead to an effective mixing between cosmic rays and target material in AGN outbursts are diffusion of cosmic rays out of the bubbles to the regions with higher ICM density \citep[e.g.][]{hinton2007}, and entrainment of non-relativistic material in the outflow from the central engine \citep[e.g.][]{pope2010}. 

\begin{figure}[ht]
\centering
\includegraphics[width=8cm]{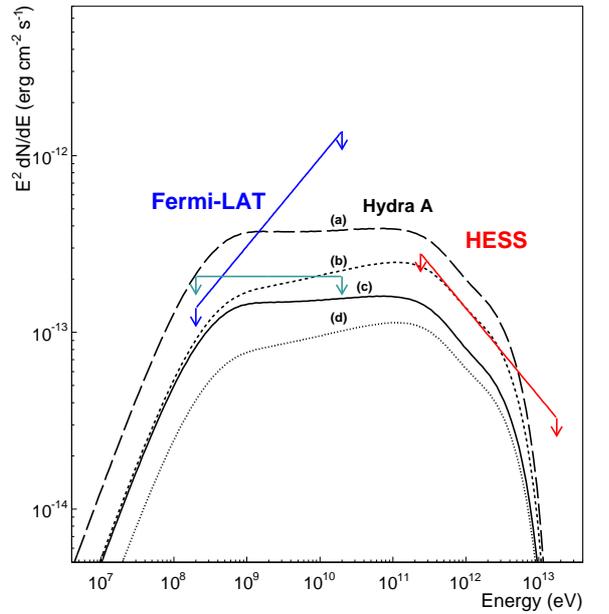}

\caption{Upper limits on the gamma-ray flux are compared to the predicted spectral energy distribution for a hadronic scenario for the Hydra~A system. \emph{Fermi} limits are shown in blue  for $\Gamma = 1.5$ and cyan for $\Gamma = 2.0$ and H.E.S.S. limits are displayed in red for $\Gamma = 2.5$. Gamma-ray indices are chosen to approximately match the shape of the predicted spectrum. Upper limits for \emph{Fermi} and H.E.S.S. for the entire probed energy range are shown with the assumed spectral index
and for consistency the same representation for the \emph{Fermi} and H.E.S.S. limits are used.
The continuation of the H.E.S.S. limit of $2.8 \times 10^{-13}$~erg~cm$^{-2}$~s$^{-1}$ above the threshold of 240~GeV towards higher energies only reflects the adopted spectral index of
$\Gamma = 2.5$ and therefore does not represent the general H.E.S.S. sensitivity at higher energies.
For details of the different models see main text.
Model (a) would predict an integral flux of $F(>240\, \mathrm{GeV}) \approx 1.5 \times 10^{-12}$~cm$^{-2}$s$^{-1}$ above the H.E.S.S. threshold.} 

\label{figure:sed}
\end{figure}

For Hydra~A order of magnitude estimates give a gamma-ray luminosity of $L_{\gamma} = E_\mathrm{pp}/3 \tau_{pp} \sim 10^{43}$~erg~s$^{-1}$, with $E_\mathrm{pp} \sim 10^{61}$~erg is the total energy in hadronic cosmic rays and $\tau_{pp} \sim 6 \times 10^{9}$~years is the cooling time for proton-proton interactions for a mean density of target material of $5 \times 10^{-3}$~cm$^{-3}$, as obtained from X-ray measurements \citep{nulsen2005b}. This results in a gamma-ray flux at Earth of $F_\gamma = L_{\gamma}/4 \pi d_L^2 \sim 10^{-12}$~erg~cm$^{-2}$~s$^{-1}$. This estimate shows that for an optimistic scenario where cosmic rays are well mixed with the embedding target material, this object is within reach of the current generation of gamma-ray instruments. 
The predicted gamma-ray flux for a more elaborate model for Hydra~A, assuming a hadronic emission mechanism and corrected for absorption by the EBL for a redshift of 0.0538, is shown for different scenarios \citep[models from][]{hinton2007} in Fig.~\ref{figure:sed}. For all cases the adopted mean density of thermal plasma outside the bubbles is $5 \times 10^{-3}$ cm$^{-3}$.

Cases (a) and (b) consider an energy in cosmic rays corresponding to the energetics of the blast wave surrounding the bubbles of $9 \times 10^{60}$ erg. In case (a) the bubbles are filled with cold, unseen gas with the same density as the surroundings which could be entrained by the AGN outflow. In case (b) bubbles are completely evacuated from the thermal ICM and mixing between the thermal ICM, and the cosmic rays results solely from energy-dependent diffusion of cosmic rays to the outside medium of the bubbles. 

For the cases (c) and (d) it is adopted that the total energy in cosmic rays is $1 pV = 4 \times 10^{60}$~erg, which is necessary to prevent the cavities in the X-ray emitting gas to collapse. In case (c) the density of the ICM in the bubbles is half of the density outside the bubbles and for case (d) the same scenario comprising empty cavities as (b) is adopted. 

Predictions of these different model assumptions are compared to the upper limits on the gamma-ray emission of the Hydra~A system obtained with \emph{Fermi}-LAT and H.E.S.S. 
From Fig.~\ref{figure:sed} it is evident that these instruments are able to constrain 
the most favorable scenario for hadronic cosmic ray content and mixing of cosmic rays and thermal gas in the Hydra~A galaxy cluster. This scenario would require a complete compound between cosmic rays and ICM. This model~(a) predicts a flux of $E^2 dN/dE \approx 4 \times 10^{-13}$~erg~cm$^{-2}$~s$^{-1}$ in the range of about 1~GeV - 300~GeV. The presence of cavities in the ICM seems to argue against complete mixing between these two components. However, it has to be noted that 10\% of the entire ICM contained within a radius of 150~kpc from the cluster center has been dredged-up by the AGN outburst. This up-lifted cooler gas partially follow the location of the giant outer bubbles \citep{gitti2012}. Since these bubbles occupy 10\% of the cluster volume within a radius of 150~kpc \citep{wise2007}, significant entrainment of cool gas in the outer bubbles can be expected.
From the upper limits obtained with \emph{Fermi}-LAT (see Tab.~\ref{table:upper_limits_fermi}) and H.E.S.S. above 240~GeV of $2.8 \times 10^{-13}$~erg~cm$^{-2}$~s$^{-1}$ (assuming $\Gamma = 2.5$) limits on the degree of mixing between cosmic rays and ICM can be derived. \emph{Fermi}-LAT can constrain the degree of mixing to 0.5 and H.E.S.S. can limit the degree of mixing to less than 0.7 where 0 means no mixing and 1 defines complete mixing between the two components. It is expected that particles with higher energies diffuse faster into their surroundings and therefore also mix faster with the ambient medium. Thus both values for the limit on the degree of mixing at different energies provide interesting constraints on particle transport in the Hydra~A system.
In general, hadronic cosmic rays as the 
energetically most important feedback agent in cluster-scale AGN 
outbursts can currently not be excluded.

\subsubsection{Electronic population}
\label{sec:leptonic}

Electrons, as opposed to hadrons, can lose their energy efficiently during the evolution time of the outburst in Hydra~A of $\sim10^8$ years \citep{hinton2007}. For magnetic fields found in Hydra A of about 6 $\mu$G \citep{taylor1993} only electrons with energies $<$~1~GeV are still present after synchrotron-cooling on such a time-scale. This picture is supported by a steepening of the radio index of synchrotron radiation from about $\alpha \simeq 0.7$ close to the core to almost $\alpha \simeq 2$ at the edge of the lobes in the radio band of 330 -- 1415 MHz \citep{lane2004}. For such a cooling-dominated electron population no gamma-ray inverse-Compton (IC) emission is expected in the Fermi-LAT energy range. 

However, \emph{Fermi}-LAT has detected gamma-ray emission from the giant lobes of the Centaurus~A system \citep{abdo2010a}. 
Since this is the only system where gamma-ray emission from an AGN outburst has been detected on spatial scales of the order of 100~kpc, it is the only example which can in principle be compared to the limits obtained for the Hydra~A lobes.
It has to be noted that Centaurus~A and Hydra~A are quite different systems. In contrast to Centaurus~A, Hydra~A is located in a galaxy cluster environment with buoyantly rising bubbles. This fact together with different jet power, differing black hole mass and different accretion history in both systems may limit the applicability of such a comparison.
 
The \emph{Fermi}-LAT discovery of the Centaurus~A lobes has been interpreted in the framework of a leptonic scenario where energetic electrons up-scatter CMB and EBL photons to gamma-ray energies. The limited radiative lifetimes of these energetic electrons may point towards in situ particle acceleration in the lobes. These processes may also be at work in the Hydra~A system. 

\begin{figure}[ht]
\centering
\includegraphics[width=9.3cm]{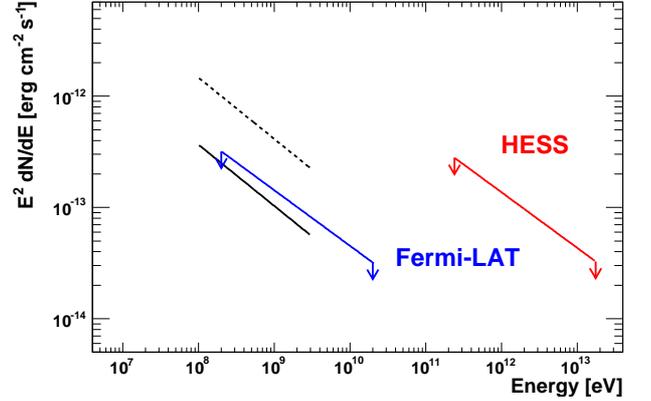}

\caption{Comparison between scaled gamma-ray flux of the Centaurus~A lobes and the upper limits for Hydra~A. \emph{Fermi} and H.E.S.S. limits are shown for $\Gamma = 2.5$. Gamma-ray indices are again chosen to approximately match the shape of the predicted spectrum. Here it is assumed that the Hydra~A lobes contain electrons with the same spectral characteristics as the Centaurus~A lobes and that CMB and EBL are the dominating photon fields which are up-scattered to gamma-ray energies. The gamma-ray flux of the Centaurus~A lobes is scaled to a distance of 240 Mpc and to a total energy in electrons of $10^{60}$ erg (solid line) and $1 pV = 4 \times 10^{60}$ erg (dotted line), respectively, to account for the different distance and energetics of the Hydra~A system.} 

\label{figure:cena}
\end{figure}

To test a similar scenario for Hydra~A as has been found for Centaurus~A, the upper limits on gamma-ray emission obtained from Hydra~A are compared to the flux measurements of the Centaurus~A lobes scaled for the different distance and energetics of the Hydra~A system. For this comparison for the combined emission of both Centaurus~A lobes a flux above 100~MeV of $1.86 \times 10^{-7}$~ph\,~cm$^{-2}$s$^{-1}$ and a spectral index of 2.55 is adopted \citep{abdo2010a}. It has to be noted that in addition to the lobes, the nucleus of Centaurus~A also emits gamma-rays with a comparable flux \citep{abdo2010b} but for the following discussion only the emission from the lobes is considered. To constrain the total energy in electrons in the lobes also the intensity of the photon field available for IC up-scattering has to be known. For the case of Centaurus~A it was found that CMB and EBL are the dominating photon fields and a total energy of electrons of $1.5 \times 10^{58}$ erg was estimated \citep{abdo2010a}. The intensities of the CMB and EBL photon fields are equivalent for the Hydra~A and the Centaurus~A systems. 
With the assumption that the Hydra~A lobes contain electrons with the same spectral characteristics as the Centaurus~A lobes and corrected for the distance to Hydra~A it is found that the total energy in electrons in the Hydra~A lobes can be constrained to $\lesssim 2 \times 10^{60}$ erg with an uncertainty dominated by the measured gamma-ray flux of the Centaurus~A lobes of about 30\% \citep{abdo2010a}. This is smaller than the 1 $pV$ work of $ 4 \times 10^{60}$ erg (see Fig.~\ref{figure:cena}). This limit can be regarded as conservative since the central galaxy in Hydra~A with log$(L_\mathrm{V}/ [\mathrm{erg/s}]) = 45.16$
(apparent V-band magnitude $m_\mathrm{V} = 12.63$, extinction $A_\mathrm{V} = 0.139$)\footnote{\url{http://ned.ipac.caltech.edu/}} is about 20 times more luminous than Centaurus~A and can therefore provide an additional photon field for IC up-scattering (see Sec.~\ref{sec:sync-ic}). 

To summarize, for an electron population with the same spectral characteristics as it is found in the Centaurus~A lobes, electrons can be excluded to be the main pressure support of the large scale bubbles in Hydra~A. This result together with the fact that radio emission from the Centaurus~A giant lobes extends to higher frequency than in in the Hydra~A giant lobes \citep{abdo2010a,lane2004} points towards quite distinct properties of the electron population in these systems. 

\subsection{Leptonic scenario for the inner lobes}
\label{sec:sync-ic}

The situation for leptonic gamma-ray emission may be more promising in the youngest, innermost lobes. Indeed the Hydra~A system is detected in the radio band up to frequencies of 90~GHz \citep{cotton2009,wright2009}. These observations show the presence of highly-energetic electrons which are relevant for IC gamma-ray production. Thus, here the upper limits on the gamma-ray emission for Hydra~A are compared to the expected gamma-ray flux from the inner radio lobes.

For estimating the expected gamma-ray IC luminosity in a leptonic scenario, first the population of electrons in the lobes is explored according to their radio synchrotron emission. The particle populations are evaluated separately for the inner and outer lobes of Hydra~A. The inner lobes \citep[denoted as `A' and `B' in][]{wise2007} are assumed to be prolate spheroids located at the distance of $r_{\rm in} = 25$\,kpc from the center, with semi-minor axis $a_{\rm in} =20''$ and semi-major axis $b_{\rm in}=35''$ \citep[as inferred from 1.4\,GHz radio maps in][]{birzan2008}. The total volume of the inner lobes is therefore $V_{\rm in} = 2 \times (4/3) \pi a_{\rm in}^2 b_{\rm in}$. For the outer lobes \citep[denoted as `E' and `F' in][]{wise2007}~ $r_{\rm out} = 225$\,kpc, $a_{\rm out}=90''$ and $b_{\rm out}=120''$ \citep{birzan2008} are used.

\begin{figure}
\includegraphics[width=9.7cm]{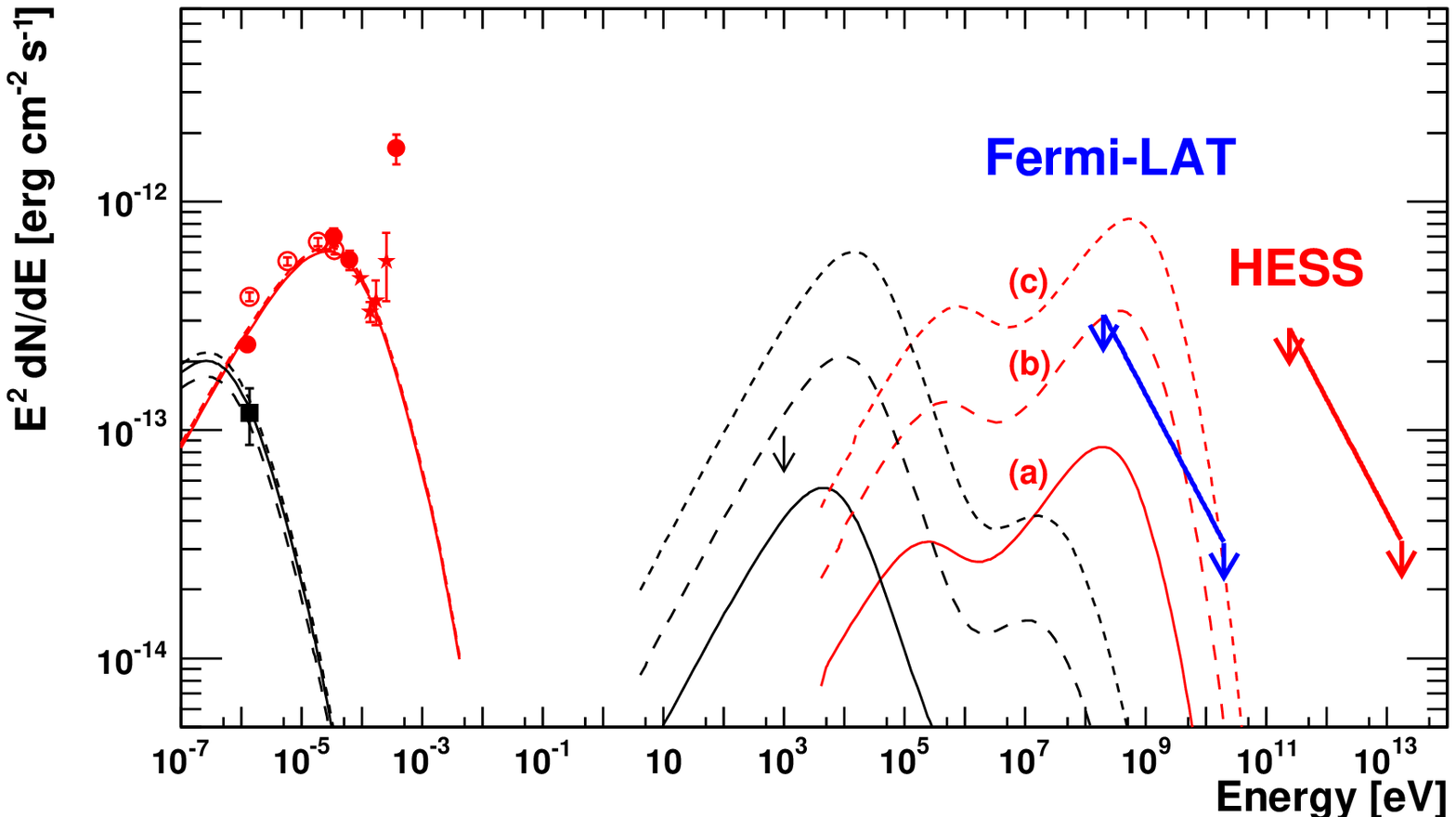}
\caption{Synchrotron and expected corresponding IC emission for the inner and outer lobes of Hydra~A. The experimental data for the inner lobes are taken from \citet{birzan2008}, \citet{cotton2009}, and \citet{wright2009} and are shown as red open circles, red filled circles, and red stars, respectively. The radio flux for the outer lobes (indicated by black squares) is adopted from \citet{birzan2008}. Additionally the upper limit for the power-law emission at 1\,keV for the outer (Northern) lobe \citep[black arrow;][]{hardcastle2010} is included. The red curves correspond to the emission of the inner lobes, and the black curves represent the outer lobes. The solid curves illustrate model (a), the dashed curve model (b) and the dotted curve is for model (c) from Tab. \ref{tab:models}. \emph{Fermi} and H.E.S.S. limits are shown for $\Gamma = 2.5$.}
\label{F-SED}
\end{figure}

The radio fluxes for the inner lobes are taken from \citet{birzan2008}, \citet{cotton2009}, and \citet{wright2009} and are shown as red open circles, red filled circles, and red stars in Fig.~\ref{F-SED}, respectively. The radio flux for the outer lobes is adopted from \citet[black square in Fig.~\ref{F-SED}]{birzan2008}. Note that the WMAP fluxes as given in \citet{wright2009} match well the radio continuum of the inner lobes, with the exception of the 61\,GHz flux. Therefore it is assumed that these fluxes represent indeed the high-energy tail of the synchrotron emission of the inner lobes. For the 61\,GHz flux as well as the 90~GHz flux \citep{cotton2009} it is anticipated that they are dominated by the flat-spectrum synchrotron emission of the jets and the nucleus of the radio galaxy in Hydra~A instead of the inner lobes.

As discussed in \citet{birzan2008} the synchrotron continua of both the inner and the outer lobes of Hydra\,A can be well represented by broken power laws with the low- and high-frequency radio spectral indices $\alpha_\mathrm{low} \simeq 0.5$, and $\alpha_\mathrm{high} \simeq 1.5$, respectively. Therefore the lobes' electron energy distribution is assumed to be $n_e(\gamma) \propto \gamma^{-2}$ for $1 \leq \gamma \leq \gamma_{\rm br}$, and $n_e(\gamma) \propto \gamma^{-4} \times \exp[-\gamma/\gamma_{\rm max}]$ for $ \gamma \geq \gamma_{\rm br}$. In the case of the inner lobes, the break and the maximum electron Lorentz factors are assumed to correspond to the synchrotron frequencies 10\,GHz and 100\,GHz, respectively. In the case of the outer lobes the analogous synchrotron break and maximum frequencies are taken as $0.1$\,GHz and 1\,GHz, respectively. 

For the evaluation of the IC-fluxes the following target photon fields are taken into account: CMB, EBL, IR emission of the nuclear dust, and the starlight emission of the elliptical host. The EBL is approximated by the spectrum given in \citet{raue2008}. For the circumnuclear dust emission a modified black body spectrum with the dust temperature $60$\,K and the total IR luminosity integrated over the frequency range $10^{11}-10^{13}$\,Hz equal to $L_{\rm dust} \simeq 7 \times 10^{43}$\,erg\,s$^{-1}$ is assumed. This model can account well for the SCUBA and MIPS data for the radio galaxy in Hydra\,A \citep{shi2005,zemcov2007,dicken2008}. The IR energy density at the position of the inner and outer lobes is therefore taken as $U_{\rm dust} = L_{\rm dust} / 4 \pi r^2 c$. In the case of the starlight emission of the host galaxy, the template spectrum as discussed in \citet{stawarz2006}, normalized to the total V-band luminosity $L_V = 1.45 \times 10^{45}$\,erg\,s$^{-1}$, is adopted. The energy density of the starlight at the position of the lobes is then evaluated as $U_{\rm star} = L_{\rm star} / 4 \pi r^2 c$. For the inner lobes the dominating photon fields are CMB in the microwave regime, the dust emission in the infrared and the star light in the optical, respectively. EBL has been found to be unimportant for IC up-scattering in the inner lobes.

With all the model parameters and assumptions as specified above, the synchrotron and inverse-Compton fluxes of the inner and outer lobes of Hydra\,A are evaluated for the remaining two free parameters: the lobes' magnetic field intensity $B$, and the equipartition ratio $\eta$. The latter parameter is defined as $\eta \equiv U_\mathrm{e}/U_\mathrm{B}$ where $U_\mathrm{e} \equiv \int d\gamma \, \gamma m_\mathrm{e} c^2 \, n_\mathrm{e}(\gamma)$, and $U_\mathrm{B} \equiv B^2 / 8 \pi$. 
These values also define the total energy in ultrarelativistic electrons and the magnetic field $E_\mathrm{e+B} = V \times (U_\mathrm{B} + U_\mathrm{e}) = V \times U_\mathrm{B} \times (1 + \eta)$ stored in the bubbles with volume $V$.
The calculations are done for different sets of the values of $B$ and $\eta$ to match the observed radio fluxes in all the cases. The results are presented in Fig.~\ref{F-SED}. There the red curves correspond to the emission of the inner lobes, and the black curves represent the outer lobes. The evaluated synchrotron fluxes (see Fig.~\ref{F-SED}) match well the collected data-set and are in general agreement with the spectral analysis carried out by \citet{lane2004} and \citet{birzan2008}. The solid curves illustrate the case with the exact electron--magnetic field energy equipartition, namely $B_{\rm in} = 16$\,$\mu$G, $\eta_{\rm in} = 1$, $B_{\rm out} = 6.3$\,$\mu$G, and $\eta_{\rm out} = 1$ (model a). Note that the standard minimum-energy calculations return typically the equipartition magnetic field $15-30$\,$\mu$G for the inner lobes and $3-6$\,$\mu$G for the outer lobes \citep[e.g.,][]{taylor1990,birzan2008} which is in agreement with our modeling. 
Parameters of further models can be found in Tab.~\ref{tab:models}. Model (b) is for a case with the magnetic field twice lower than the equipartition value, and in model (c) the magnetic field is three times below the equipartition value.

\begin{table}[h]
\begin{center}

\vspace{0.2cm}
\begin{tabular}{lcccccc}
\hline
\hline
model & $B_{\rm in}$& $\eta_{\rm in} $ &$E_\mathrm{e+B, in}$ &$B_{\rm out}$ & $\eta_{\rm out}$ & $E_\mathrm{e+B, out}$ \\
 & [$\mu$G] & & [erg] & [$\mu$G] & & [erg] \\
\hline
 & & & & & & \\
a & 16 & 1 & $4.1\times 10^{58}$ & 6.3 & 1 & $4.3\times 10^{59}$ \\
b & 8  & 12 &$1.2\times 10^{59}$ & 3 & 12 & $1.2\times 10^{60}$ \\
c & 5  & 65 &$2.6\times 10^{59}$ & 2 & 65 & $2.9\times 10^{60}$ \\
\hline

\end{tabular}
\caption{Properties of the models to calculate the IC emission from the inner and outer lobes. $B$ gives the magnetic field inside the lobes, $\eta$ is the equipartition ratio and $E_\mathrm{e+B}$ is the total energy in electrons and magentic field for the inner (subscript in) and outer (subscript out) lobes, respectively. } \label{tab:models}
\end{center}
\end{table}

These calculations are now compared to the upper limits obtained in the gamma-ray range. Additionally also the upper limit for the power-law emission at 1\,keV for the outer (Northern) lobe \citep[black arrow in Fig. \ref{F-SED},][]{hardcastle2010} obtained with XMM is included. The inverse-Compton emission of the lobes in Hydra\,A is expected to be negligible at TeV photon energies. It may however be pronounced within the lower energy range from keV--to--GeV photons. The XMM and {\it Fermi}-LAT upper limits for the Hydra\,A system seem to already exclude a magnetic field in the lobes below half of the equipartition value.

\section{Conclusion}

In this paper the nearby galaxy cluster Hydra~A that hosts a cluster-scale AGN outburst is investigated for being a gamma-ray emitter. Galaxy clusters hosting a cluster-scale AGN outburst are potentially detectable gamma-ray sources due to the enormous energetics inferred from the observed AGN - ICM interactions. However, only upper limits could be obtained from 20.2 hours of H.E.S.S. and 38 month of \emph{Fermi}-LAT observations.
The non-detection of Hydra~A in gamma-rays has important implications on the particle populations and physical conditions inside the bubbles in this system.
These upper limits constrain the total energy contained in relativistic particles such as hadronic cosmic rays and electrons, which can be compared to the energy which is necessary to prevent the observed cavities in the ICM from collapsing.

Constraints on the particle population in the Hydra~A galaxy cluster can also be compared to the limits on such a non-thermal component inferred for the Perseus cluster \citep{aleksic2012b}. The Perseus cluster is an interesting candidate for this comparison since it also shows signatures of AGN -- ICM interactions in form of radio lobes and cavities in the ICM \citep[e.g.][]{fabian2011}. For the Perseus cluster \citet{aleksic2012b} constrained the average fraction of energy in hadronic cosmic rays to thermal energy $E_\mathrm{CR}/E_\mathrm{th}$ to $\lesssim 1-2$\% depending on the exact assumptions. For Hydra~A the hadronic cosmic ray content in the central 200~kpc can be limited to about $5 \times 10^{60}$\,erg, assuming complete mixing between cosmic rays and ICM (see Fig. \ref{figure:sed}). When adopting a total gas mass of $5 \times 10^{12}$\,M$_\odot$ and a temperature of 4 keV for the central 200\,kpc of Hydra~A \citep{david2001} then $E_\mathrm{CR}/E_\mathrm{th} \lesssim 13$\% is found. This is significantly less constraining than for the case of the Perseus cluster. Hydra~A, however, is the prime candidate to explore the particle content of giant AGN-blown lobes in galaxy clusters. This follows from the fact that the AGN outburst in the Perseus cluster is about an order of magnitude less energetic \citep[$pV \approx 3 \times 10^{59}$\,erg,][]{fabian2011} than the AGN feedback in Hydra~A. The smaller energetics is not readily compensated by the shorter distance to the Perseus cluster ($d_\mathrm{L}$ = 75~Mpc). 

For Hydra~A for the case of bubbles mainly supported by hadronic cosmic rays these upper limits can exclude the 
most favorable model, that requires full mixing between relativistic particles and embedding thermal medium.
It is found that \emph{Fermi}-LAT can constrain the degree of mixing to 50\% and H.E.S.S. can limit the degree of mixing to less than 70\%. However, hadronic cosmic rays still
remain a viable pressure support for the bubbles. 

In contrast to hadrons, electrons cool quite fast above GeV energies in the environment of the Hydra~A system. Consequently, a passively evolving population of electrons in the oldest outer lobes cannot be detected with the presented observations. However, for the youngest, inner radio lobes, the limit on the IC flux seems to exclude a magnetic field below about 8 \,$\mu$G, that is half of the equipartition value. For the large outer lobes, a population of electrons rejuvenated by in situ particle acceleration comparable to the one detected in the Centaurus~A system can be excluded as the main pressure support of the bubbles. Upper limits in the VHE gamma-ray range are not constraining for leptonic scenarios with respect to limits obtained in the GeV range.

The main feedback agent which drives the evolution of the cavities in the ICM in the Hydra~A galaxy cluster still remains unidentified. The upcoming Cherenkov Telescope Array \citep[CTA,][]{actis2011} with its increased sensitivity will be crucial to test especially the presence of hadronic cosmic rays in the Hydra~A galaxy cluster.

\acknowledgements{
The support of the Namibian authorities and of the University of Namibia
in facilitating the construction and operation of H.E.S.S. is gratefully
acknowledged, as is the support by the German Ministry for Education and
Research (BMBF), the Max Planck Society, the French Ministry for Research,
the CNRS-IN2P3 and the Astroparticle Interdisciplinary Programme of the
CNRS, the U.K. Science and Technology Facilities Council (STFC),
the IPNP of the Charles University, the Polish Ministry of Science and 
Higher Education, the South African Department of
Science and Technology and National Research Foundation, and by the
University of Namibia. We appreciate the excellent work of the technical
support staff in Berlin, Durham, Hamburg, Heidelberg, Palaiseau, Paris,
Saclay, and in Namibia in the construction and operation of the
equipment.
}

\end{document}